\documentclass[
reprint,
superscriptaddress,
 amsmath,amssymb,
 aps,
 prl,
]{revtex4-2}

\usepackage{graphicx}
\usepackage{array}
\usepackage{tabularx}
\usepackage{dcolumn}
\usepackage{bm}
\usepackage{hyperref}
\hypersetup{
    colorlinks=true,
    linkcolor=blue,
    filecolor=magenta,      
    urlcolor=cyan,
    }
\usepackage{xcolor}
\usepackage{siunitx}
\begin{document}

\title{Microplankton life histories revealed by holographic microscopy and deep learning}

\author{Harshith Bachimanchi}
\affiliation{Department of Physics, University of Gothenburg, Sweden}

\author{Benjamin Midtvedt}
\affiliation{Department of Physics, University of Gothenburg, Sweden}

\author{Daniel Midtvedt}
\affiliation{Department of Physics, University of Gothenburg, Sweden}

\author{Erik Selander}
\affiliation{Department of Marine Sciences, University of Gothenburg, Sweden}

\author{Giovanni Volpe}
\email{giovanni.volpe@physics.gu.se}
\affiliation{Department of Physics, University of Gothenburg, Sweden}

\date{\today}

\begin{abstract}
The marine microbial food web plays a central role in the global carbon cycle.
Our mechanistic understanding of the ocean, however, is biased towards its larger constituents, while rates and biomass fluxes in the microbial food web are mainly inferred from indirect measurements and ensemble averages.
Yet, resolution at the level of the individual microplankton is required to advance our understanding of the oceanic food web.
Here, we demonstrate that, by combining holographic microscopy with deep learning, we can follow microplanktons throughout their lifespan, continuously measuring their three dimensional position and dry mass. 
The deep learning algorithms circumvent the computationally intensive processing of holographic data and allow rapid measurements over extended time periods.
This permits us to reliably estimate growth rates, both in terms of dry mass increase and cell divisions, as well as to measure trophic interactions between species such as predation events. The individual resolution provides information about selectivity, individual feeding rates and handling times for individual microplanktons. 
This method is particularly useful to explore the flux of carbon through micro-zooplankton, the most important and least known group of primary consumers in the global oceans. 
We exemplify this by detailed descriptions of micro-zooplankton feeding events, cell divisions, and long term monitoring of single cells from division to division.
\end{abstract}

\maketitle

The role of herbivores in structuring plant communities is well established in terrestrial ecology. Already Darwin, in his foundations on evolutionary biology \cite{darwin1859origin}, noted how excluding herbivores from a heath land transformed it into a forest of pine trees with an altogether different species composition. 
Single celled micro-zooplankton take on the role of herbivores in the ocean, consuming approximately two thirds (\SI{40}{\peta\gram} carbon) of the primary production \cite{calbet2004phytoplankton}. 
In oceanic ecology, the primary production is dominated by unicellular phytoplankton, which produce around $65\,{\rm Pg}$ of carbon annually, quantitatively slightly exceeding the production of terrestrial plants \cite{behrenfeld1997photosynthetic,field1998primary}.
Selective grazing shapes the plankton community and drives large-scale processes such as harmful algal bloom formation and carbon export \cite{irigoien2005phytoplankton,selander2019copepods}.

Despite its importance, our understanding of the role of micro-zooplankton in shaping oceanic communities is still much less developed than that of macro-organisms that can more readily be observed at the individual level \cite{glibert2022webs}.
In fact, rates and fluxes in the oceanic microbial food web are still mainly inferred from indirect measurements or ensemble averages, leaving us with a limited mechanistic understanding. 
Quantitative estimates of primary production are mostly inferred from satellite images of ocean color (chlorophyll) using moderate resolution spectro-radiometers calibrated against in-situ isotope incorporation experiments \cite{hu2012chlorophyll}. 
Ensemble-level biomass transitions during grazing events by microscopic zooplankton are calculated from dilution experiments \cite{landry1982estimating}, where the grazer density is manipulated by dilution, and the corresponding net increase in primary production is approximated.
While these methods provide good estimates of the magnitude of biomass fluxes, they do not resolve the small-scale individual interactions that drive the large-scale processes.
Moreover, indirect measurements of processes such as micro-zooplankton grazing rest on assumptions that are not always fulfilled.
For example, feeding rates and growth rates of both predators and prey need to be unaffected by dilution, which is often not true \cite{dolan2000dilution}. In addition, the dilution technique is based on chlorophyll measurements and does not account for consumption of non-chlorophyll-bearing particles which leads to underestimation of carbon transfer \cite{stoecker2017mixotrophy}. 

Currently, the biomass of individuals is often inferred from ``volume-to-carbon'' relationships developed over time for different trophic groups of planktons \cite{strathmann1967estimating,menden2000carbon}, which require cell counting and sizing followed by elemental analysis, but do not allow continuous measurements of the same individual. 
However, these regression relations are not very precise: the average deviation of individual data points to the regressed expression exceeds 50\% \cite{menden2000carbon}. 
In addition, single cells of the same volume can differ by a factor two in dry mass, which is not possible to detect by ``volume-to-carbon'' relationships.
To go beyond the current level of detail in marine microbial food webs, we need complementary techniques that can follow individual microplanktons over extended periods, while continuously monitoring their growth rate and predation events. 

Continuous measurements can be realized using microscopy techniques. 
For example, holographic microscopy can record holograms of cells under investigation in the form of interference patterns containing phase and amplitude information.
The information in the holograms can be used to extract the three-dimensional position of microplanktons as well as their mass \cite{zangle2014live}.
Holographic imaging has already found applications in microbial, especially for in-situ measurements of particle size distributions and their identity \cite{nayakreview}.
However, its full potential has not yet been exploited, namely for the quantitative investigations of the growth and feeding patterns of individual planktons over prolonged times. 
Arguably, this is because the data acquisition and processing pipelines are very computationally expensive.

Here, we solve this problem by employing a technique that combines holography with deep learning.
The deep learning algorithms circumvents the long computational times and, once trained, allow rapid determination of three-dimensional position and dry mass of individual microplanktons over extended time periods.
We evaluate this method on nine plankton species belonging to different trophic levels and representing the major classes of microplankton.
We highlight that unlike other methods, our approach makes it possible to follow and weigh single cells throughout their lifetime, being especially useful to detail micro-zooplankton and mixotrophic life histories and feeding events. 
Furthermore, the estimated dry mass can be tagged to single planktons detected in the experimental images.
We can track and identify both prey and predator cells and closely follow the transfer of mass from cell to cell.
Finally, we observe the growth and cell divisions in diatoms by long-term monitoring of the single-cells over more than one cell cycle.

\begin{figure*}[ht]
    \includegraphics[scale=0.87, angle=0]{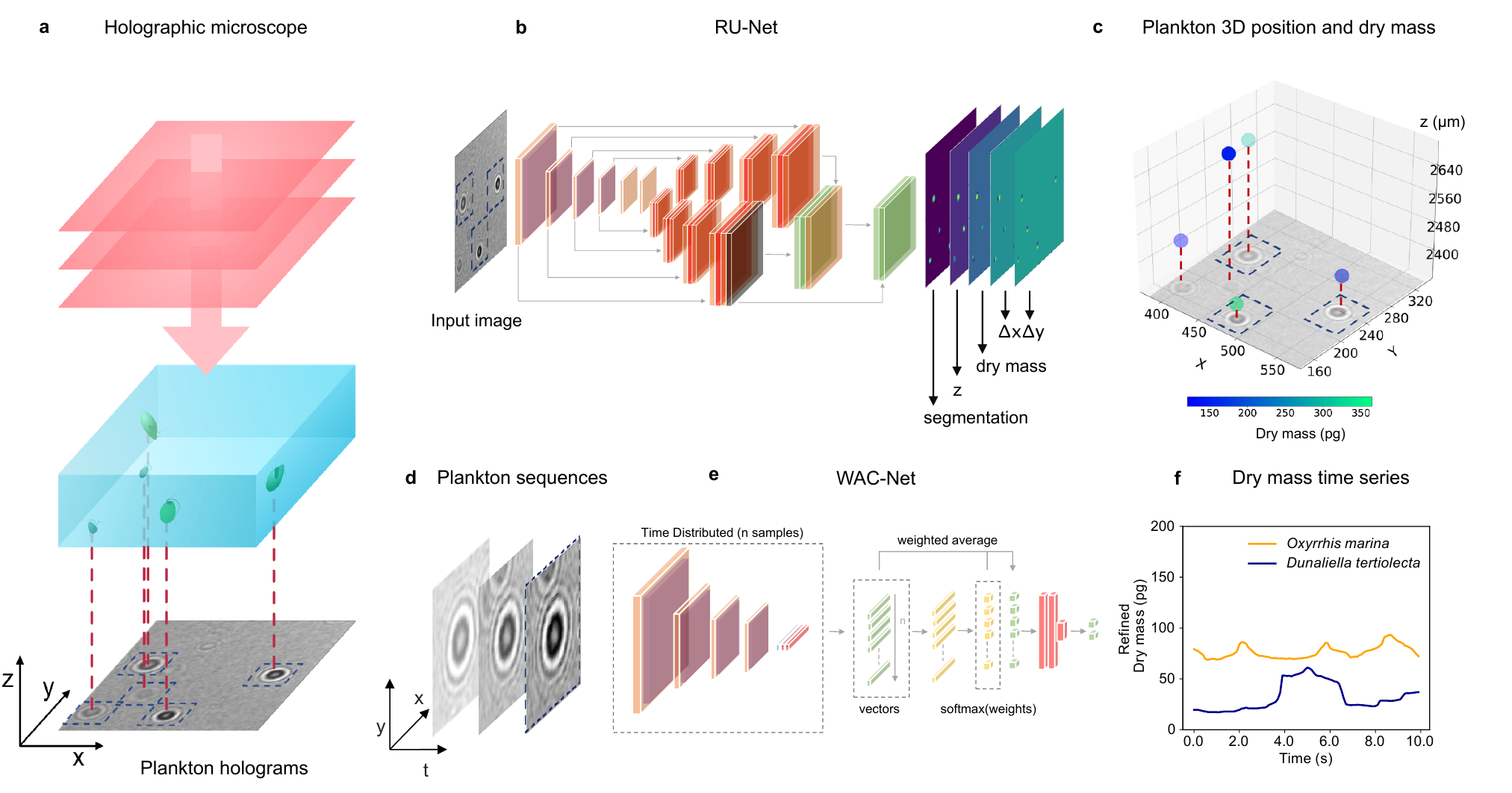}
    \caption{
    \label{figure1} 
    \textbf{Experimental setup and deep-learning data analysis.} 
    {\bf a} Holographic microscope: Planktons suspended in a miniature sample well are imaged with an inline holographic microscope. The (cropped) example holographic image features two different plankton species: \textit{Oxyrrhis marina} and \textit{Dunaliella tertiolecta} (full image in Suppl. Fig.~\ref{SF1}). 
    {\bf b} Deep-learning network 1: A regression U-Net (RU-Net, see details in Suppl. Fig.~\ref{SF2}), trained on simulated holograms, uses individual holograms to predict output maps containing the segmentation of the planktons, their $z$-position, their dry mass $m$, and the distances $\Delta x$ and $\Delta y$ from the closest plankton for each pixel (to be used for the accurate localization of planktons).
    {\bf c} Plankton 3D position and dry mass: The information obtained by the RU-Net permits us to reconstruct the 3D position of the planktons along with their dry mass (color bar). 
    {\bf d} Plankton sequences: Using the plankton positions obtained by the RU-Net, we extract sequences of $64\times64$-pixel holograms centered on an individual plankton.
    {\bf e} Deep-learning network 2: The sequences are then used by a weighted-average convolutional neural network (WAC-Net, see details in Suppl. Fig.~\ref{SF3}), trained on simulated data, to refine the estimations of $m$ and $z$. 
    {\bf f} Dry mass time series: Example of a refined dry mass prediction for a micro-zooplankton ({\it Oxyrrhis marina}, orange line) and a phytoplankton ({\it Dunaliella tertiolecta}, blue line) obtained by the WAC-Net.}
\end{figure*}

\section{Results}

{\bf Experimental setup and deep-learning data analysis.}
Fig.~\ref{figure1} shows an overview of the holographic microscopy experimental setup and the deep-learning data analysis pipeline to estimate the position and dry mass of the planktons. 
We use an inline holographic microscope in a lens-less configuration 
(see details in Methods, ``Holographic imaging''). 
A monochromatic LED light source illuminates the sample well (cylindrical well, diameter: $3\,{\rm mm}$, height: $1\,{\rm mm}$) that contains a suspension with the planktons under investigation. 
As the light passes through the well, it acquires a complex amplitude that depends on the optical properties of the materials it traverses, generating inline holograms (Suppl. Fig.~\ref{SF1}), which encode the three-dimensional position of the planktons as well as their size and refractive index.
A CMOS camera located on the opposite side of the sample well acquires the holograms for further analysis with a frame rate of $10\,{\rm fps}$, and an exposure time of $8\,{\rm ms}$.

In order to measure the position and dry mass of the planktons, the recorded holograms are analyzed by a regression U-Net (RU-Net, Fig.~\ref{figure1}b and Suppl. Fig.~\ref{SF2}, see details in Methods, ``RU-Net architecture and training'').
The RU-Net is a deep-learning architecture based on a modified U-Net, with two parallel arms in the upsampling path.
The output of the RU-Net is a five-channel image where each channel corresponds to a heat map containing: a segmentation of the planktons from the background used to obtain a rough estimate of their $xy$ (in-plane) position; their estimated $z$ (axial) position; the plankton estimated dry mass $m$; and the distances $\Delta x$ and $\Delta y$ from the closest plankton for each pixel (used to improve the in-plane localization). 
This RU-Net is implemented and trained on simulated input--output image pairs (4000 samples) using the Python software package DeepTrack 2.0 \cite{midtvedt2021quantitative}.
The output heat maps are finally processed to obtain a prediction of the plankton three-dimensional position and of their dry mass, as shown in Fig.~\ref{figure1}c. 

In order to increase the accuracy of the estimations of dry mass, $m$, we extract time sequences of holographic images cropped around an individual plankton (Fig.~\ref{figure1}d and Suppl. Fig.~\ref{SF3}) and further analyze them with a second deep-learning network.
This is a weighted-average convolutional neural network (WAC-Net \cite{midtvedt2021fast}, Fig.~\ref{figure1}e and Suppl. Fig.~\ref{SF3}, see details in Methods, ``WAC-Net architecture and training'').
The WAC-Net determines a single estimated value of the equivalent spherical radius (ESR), as well as a more accurate value of the dry mass of the plankton in the sequence, through a weighted average of the latent representation of various holograms, where the weights are learnable. The number of frames in the sequence is limited to $15$ frames for training the WAC-Net. For inference, the length of the sequence is dependent on the application. For example, when analyzing feeding events we aim to capture dry mass dynamics on short time scales, and the sequence length is therefore restricted to a single frame. For the division events, the sequence length is $15$ frames, as they occur over longer times ranging from hours to days with more number of recorded frames.
Also the WAC-Net is implemented and trained with simulated data (4000 15-frame sequences of $64\,{\rm px} \times 64\,{\rm px}$ images) using DeepTrack 2.0 \cite{midtvedt2021quantitative}.
Fig.~\ref{figure1}f shows an example of the output of the WAC-Net when applied on a sliding window over a sequence of holograms corresponding to a micro-zooplankton ({\it Oxyrrhis marina}) and a phytoplankton ({\it Dunaliella tertiolecta}).

\begin{figure*}[pt]
    \includegraphics[scale=0.93, angle=0]{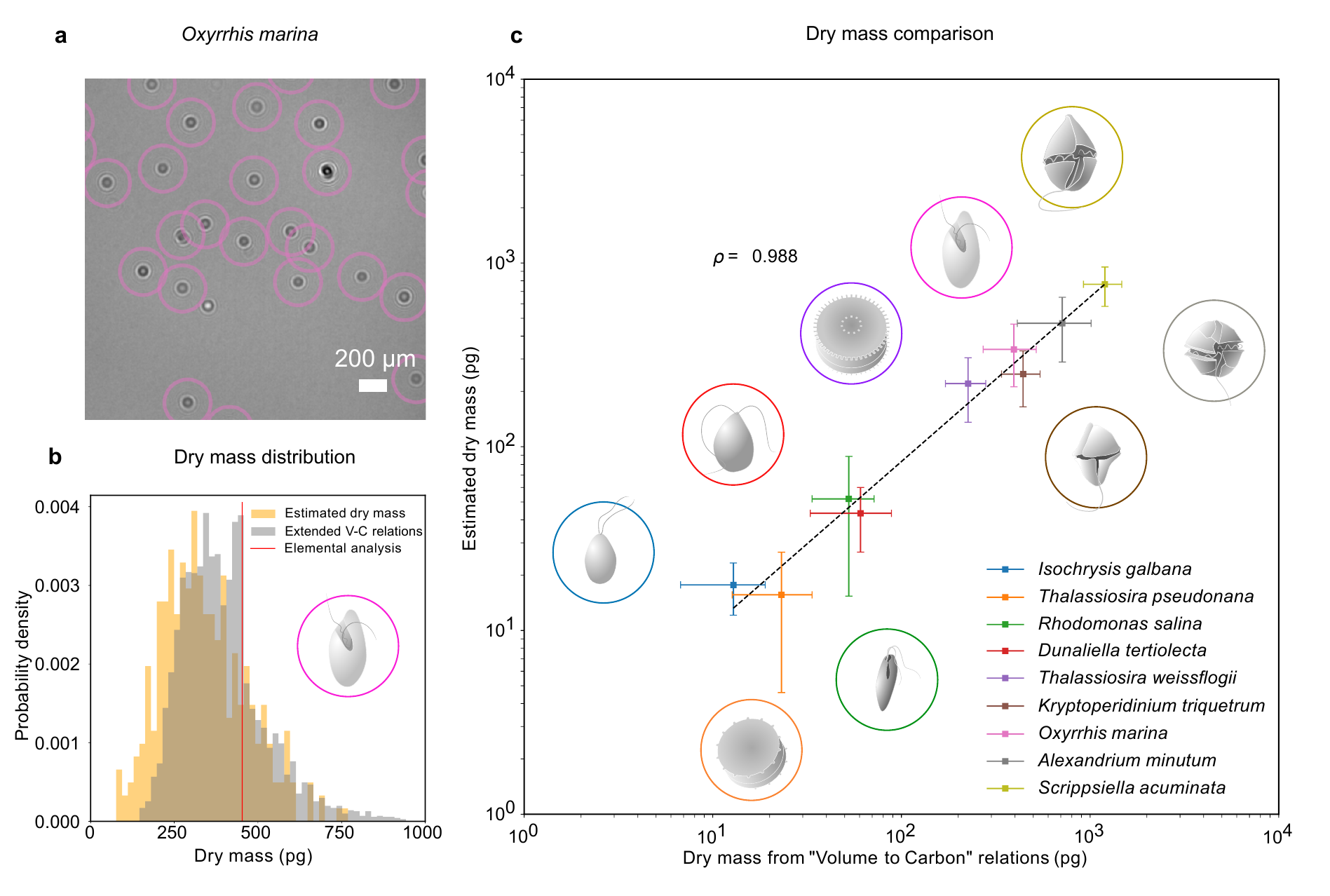}
    \caption{
    \label{figure2} 
    {\bf Dry mass estimates.} 
    {\bf a} Phytoplankton species \textit{Oxyrrhis marina} as detected by RU-Net on a portion of experimental hologram (see Suppl. Fig.~\ref{SF1} for the complete hologram). 
    {\bf b} Dry mass distributions for \textit{O. marina} (illustrated in the inset) obtained by applying WAC-Net to the experimental holograms (orange) and by ``volume-to-carbon'' relationships (gray, \cite{menden2000carbon}).
    The red line is the value of the average mass estimate obtained from elemental analysis. 
    {\bf c} Comparison of the dry mass estimations obtained by WAC-Net and by the ``volume-to-carbon'' method for nine different species of diatoms (\textit{Thalassiosira pseudonana}, \textit{Thalassiosira weissflogii}), phytoplantons (\textit{Isochrysis galbana}, \textit{Rhodomonas salina}, \textit{D. tertiolecta}), and micro-zooplanktons (\textit{Oxyrrhis marina}, \textit{Kryptoperidinium triquetrum}, \textit{Alexandrium minutum}, \textit{Scrippsiella acuminata}). 
    The two measurements have a correlation coefficient of $\rho = 0.988$. The dashed line represents the best fit and the error bars show the standard deviation of the distributions.
    The insets illustrate each species.
    }
\end{figure*}

{\bf Dry mass estimates.}
The combination of RU-Net and WAC-Net permits us to measure the dry mass of each plankton at any point in time. 
For example, Fig.~\ref{figure2}a shows a portion of an inline hologram of the micro-zooplankton species, \textit{Oxyrrhia marina} tracked by the RU-Net (circles). 
Individual \textit{O. marina} cells are then traced for 30 frames and their holograms are further processed with WAC-Net to obtain an estimation of the dry mass for each cell. 
The orange histogram in Fig.~\ref{figure2}b shows the dry mass distribution estimated by the WAC-Net.

To benchmark the dry mass measurements, we used the ``volume-to-carbon'' relationships from Ref.~\cite{menden2000carbon} followed by an extrapolation of elemental composition using extended Redfield ratios \cite{anderson1995hydrogen} (see Methods, ``Dry mass estimation by volume-to-carbon relationships''). 
The gray histogram in Fig.~\ref{figure2}b shows the results for the case of \textit{O. marina}.
The dry mass predicted by the ``volume-to-carbon'' relationships is obtained as $394\pm 123 \,{\rm pg}$ (the uncertainity represents the standard deviations of the distribution), which agrees well with the dry mass estimated by our technique is $338\pm 126\,{\rm pg}$ (orange histogram).
We highlight that, in contrast to the ``volume-to-carbon'' relation method, the dry mass estimated by our approach can be tagged to individual cells in the image. 
This additional feature can be used to study the dry mass evolution of single cells (e.g., in the following sections, we will exploit this possibility in two exemplary studies of feeding and cell-division events).

We then repeat this analysis for nine species of planktons belonging to different taxonomic groups and trophic levels in the marine ecosystem (See Methods, ``Plankton cultures''): phytoplankton species (\textit{Isochrysis galbana}, \textit{Rhodomonas salina}, \textit{Dunaliella tertiolecta}); micro-zooplankton species (\textit{Kryptoperidinium triquetrum}, \textit{Alexandrium minutum}, \textit{Scrippsiella acuminata}, along with \textit{Oxyrrhis marina} which is used in the above discussion); and diatomic species (\textit{Thalassiosira weissflogii}, \textit{Thalassiosira pseudonana}).
These results are summarized in Fig.~\ref{figure2}c.
The data points and error bars represent the mean and standard deviations of the dry mass distributions estimated by our method and the ``volume-to-carbon'' relations method. 
The two estimates correlate very well (correlation coefficient $\rho = 0.988$). 
A detailed dry mass distribution comparison (along with equivalent spherical radius distribution comparison) for different species can be seen in Suppl. Fig.~\ref{SF4}.

As a further independent test, we also estimated dry mass from the elemental analysis of carbon and nitrogen content in \textit{O. marina} (extrapolated to the other fundamental elements hydrogen, oxygen and phosphorous through Redfield ratios \cite{anderson1995hydrogen}). The resulting dry mass ($453\, {\rm pg}$, indicated with a red line in Fig.~\ref{figure2}) also confirms that our method arrives at realistic numbers. 
We note that the holographic method estimates the total dry mass, while the elemental analysis measures carbon and nitrogen; thus,  we assumed that other elements are present in amounts consistent with extended Redfield ratios (average phytoplankton composition \cite{anderson1995hydrogen} to estimate the dry mass from elemental analysis, see Methods, ``Dry mass estimation by elemental analysis''). 
The average value indicated by the red line in Fig.~\ref{figure2}b lies with in the distributions predicted by holographic estimate.

\begin{figure*}[pt]
    \includegraphics[scale=0.85, angle=0]{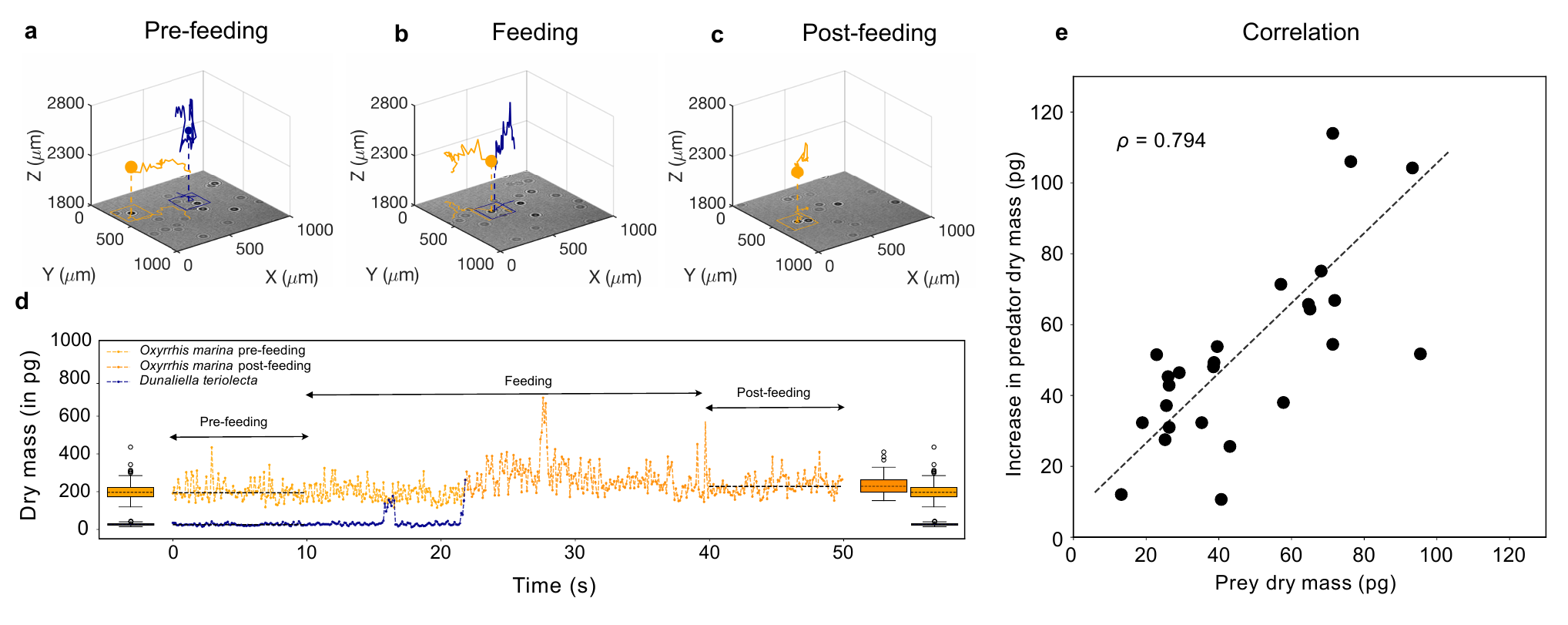}
    \caption{
    \label{figure3} 
    {\bf Feeding events.} 
    {\bf a}-{\bf c} 3D recording of a feeding event where 
    {\bf a} a predator micro-zooplankton (\textit{Oxyrrhis marina}, orange traces) approaches a prey phytoplankton (\textit{Dunaliella teriolecta}, blue traces), {\bf b} feeds on it, and {\bf c} finally moves away. 
    The 2D projection of traces is superimposed on the holographic images in the bottom (see also Suppl. Fig.~\ref{SF5}). 
    {\bf d} Dry mass time series of predator (orange trace) and prey (blue trace) estimated by WAC-Net in the three different phases. 
    The pre-feeding prey and predator dry mass distributions are represented in the box plots on the left, and the predator post-feeding one on the right (here also also the pre-feeding ones are reproduced for easier comparison):
    The post-feeding dry mass increment of the predator matches the dry mass of the prey.
    {\bf e} There is a high correlation ($\rho=0.794$) between dry mass increments of predators and dry mass of prey for 26 different feeding events.
    }
\end{figure*}

\subsection{Feeding events}

We use the phytoplankton species \textit{Dunaliella tertiolecta} and the micro-zooplankton species \textit{Oxyrrhis marina} as the prey and predator, respectively. Figs.~\ref{figure3}a-c show the 3D traces of prey (blue) and predator (orange) during a feeding event (see 3D movie of the feeding event in Suppl. Video~\ref{SV1}). 
In the pre-feeding phase (Fig.~\ref{figure3}a, corresponding to about $10\,{\rm s}$ or $100\,{\rm frames}$), the predator explores the sample volume in a random fashion. It passes the prey cell closely on a couple of occasions before it makes contact (see Suppl. Videos~\ref{SV1} and ~\ref{SV2}).
In the feeding phase (Fig.~\ref{figure3}b, about $30\,{\rm s}$ or $300\,{\rm frames}$), the predator makes contact with the prey and performs a localised swirling motion about a fixed location for 16 s while handling the prey.
In the post-feeding phase (Fig.~\ref{figure3}c, about last $10\,{\rm s}$ or $100\,{\rm frames}$), the predator returns back to its normal swimming behaviour and carries on its search for new prey. 

Fig.~\ref{figure3}d shows the dry-mass time series of prey and predator during the feeding event. 
As the feeding events happen on a short time scale compared to the frame rate of the camera, we use WAC-Net with a sliding window of only one frame, maximizing the available temporal resolution of the dry mass estimation. 
The dry mass distributions in the pre-feeding phase are shown by the box plots on the left hand side of the plot: the prey is $26\pm 1 \,{\rm pg}$ (blue box plot) and the predator $204\pm 5 \,{\rm pg}$ (orange box plot). The uncertainties represent the standard error of the mean.
The post-feeding dry mass distribution of the predator is shown on the right hand side of Fig.~\ref{figure3}d (orange box plot, for easier comparison, also the pre-feeding dry mass distributions are shown):
it is $234\pm 5 \,{\rm pg}$.
The difference in predator dry mass post- and pre-feeding closely matches the prey dry mass.
This indicates that the predator has fully consumed its prey and it provides a direct measurement of the amount of carbon consumed during each individual feeding event.

In Fig.~\ref{figure3}e, we report the results of the dry mass increase in 26 different feeding events.
The increase in the predator dry mass in the post-feeding phase correlates well with the pre-feeding dry mass of the prey (correlation coefficient $\rho = 0.794$). 
The slope of the best fit line (with slope, $\alpha = 0.97$) also indicates that on average 97\% of prey is consumed by the predator in a feeding event.
Thus it is possible to quantify individual feeding rates and, if predator cells are followed over time, also gross growth efficiency, i.e., how much of the consumed biomass is converted into predator biomass. 

\subsection{Life history of a plankton}

\begin{figure*}[pt]
    \includegraphics[scale=0.89, angle=0]{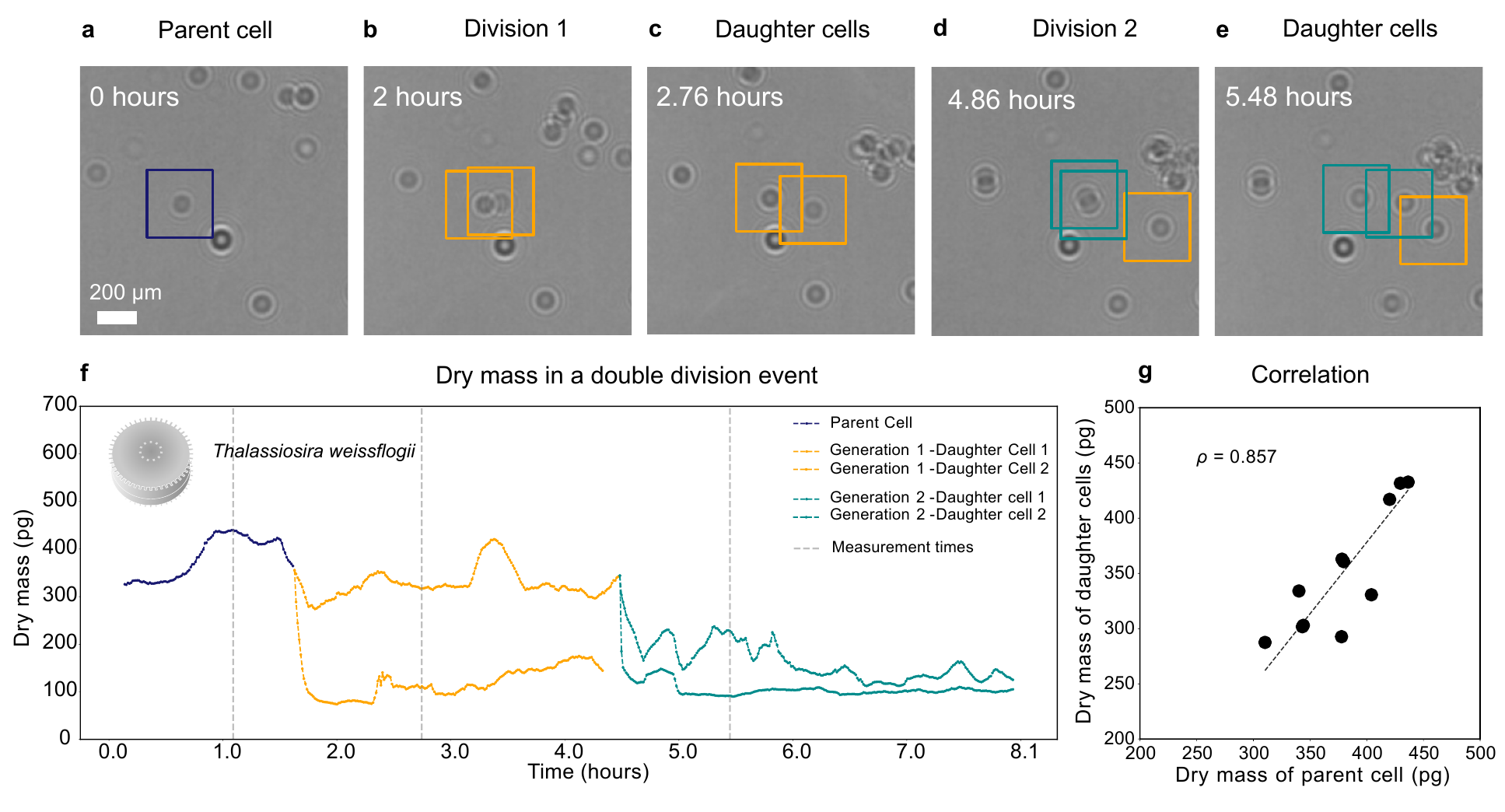}
    \centering
    \caption{
    \label{figure4} 
    {\bf Growth and cell division of a diatom.} 
    {\bf a}-{\bf e} Different life stages of a diatom (\textit{Thallasiosira weissflogii}) and its daughter cells:
    {\bf a} the parent cell (blue) 
    {\bf b} divides into two daughter cells (orange);
    {\bf c} the daughter cells continue to grow, 
    {\bf d}-{\bf e} until another cell division.
    {\bf f} Dry mass time series through generations estimated by WAC-Net (see also Suppl. video~\ref{SV3}). 
    Each cell dry mass is estimated when it has at least $3.6\,{\rm \mu m}$ ($40\,{\rm px}$) of empty space around it to ensure optimal performance of the WAC-Net; the corresponding times are indicated by the gray dashed lines.
    A drop in the dry mass values can be noticed with the daughter cells in subsequent divisions. 
    {\bf e} Correlation plot showing the relation between sum of the dry masses of the daughter cells and the dry mass of the parent cell for 11 different division events ($\rho = 0.857$).
    }
\end{figure*}

The technique we have developed can follow the entire life histories of planktons, over time scales from hours to days.
To demonstrate this, we use a diatomic species, \textit{Thalassiosira weissflogii}, which is auto-trophic and  non-motile.
Over a preriod of 8 hours (Fig.~\ref{figure4}), we image a \textit{T. weissflogii} and two generations of its daughter cells, continuously assessing the changes in their dry mass using the WAC-Net (which we already used to estimate the dry mass of \textit{T. weissflogii} in Fig.~\ref{figure2}c, (see Methods, ``Holographic imaging''). 
We place a low-density ($1000\,{\rm cells \,ml^{-1}}$) culture of diatoms in the sample well, which we  illuminate with a white light source ($5\,{\rm W}$, $60\,{\rm Hz}$ warm light source bulb, aligned not to affect the holographic imaging sensor) to aid the cell growth. 

Figs.~\ref{figure4}a-e show the growth and division of a diatom imaged over a small portion of the sample.
The parent cell (highlighted in Fig.~\ref{figure4}a) initially divides into two daughter cells, approximately 0.14 hours into the experiment (Fig.~\ref{figure4}b). Note that the biomass does not divide equally between the daughter cells. This type of asymmetric divisions have been shown in both bacteria and diatoms before, but mainly in size. Here the daughter-cells receive unequal proportions of the biomass from the mother cell. 
Then, the two daughter cells move slightly apart (Fig.~\ref{figure4}c) and the cell with the largest biomass of the two divides again at 4.86 hours (Figs.~\ref{figure4}d-e).

Fig.~\ref{figure4}f shows the dry mass of the parent and daughter cells as the experiment proceeds. 
We remark that, while the dry mass of these cells is continuously monitored, the WAC-Net estimates the most reliable values when the cells are isolated. 
Therefore, we consider the reference dry mass measurements as those when the cells have at least $3.6\,{\rm \mu m}$ ($40\,{\rm px}$) of empty space around them before or after each division; these times are indicated by the grey dashed lines in Fig.~\ref{figure4}f.
The initial parent cell dry mass (measured at 1.1 hours) is estimated at $433\pm 2\,{\rm pg}$. 
The dry mass of its two daughter cells (measured at 2.7 hours, as soon as the two daughter cells move sufficiently apart) are $326\pm 1\,{\rm pg}$ and $110\pm 1\,{\rm pg}$, whose sum is close to the dry mass of the parent cell.
As the experiment proceeds, one of the daughter cells divides again producing a second generation of daughter cells (Fig.~\ref{figure4}d), whose dry masses are $225\pm 3\,{\rm pg}$ and $93\pm 1\,{\rm pg}$ (at 5.48 hours, Fig.~\ref{figure4}e). Again, their sum is close to the mass of their parent cell.
The uncertainty in the dry mass value represents the standard error of the mean computed for {$\pm 5$} frames around the measurement point (gray dashed lines in Fig.~\ref{figure4}).

We have repeated this experiment with various cell densities with independently cultured samples, collecting multiple division events. 
Fig.~\ref{figure4}g shows the high correlation ($\rho=0.857$) between the parent cell dry mass and the sum of the daughter cells' dry masses for 11 cell divisions. 
It is interesting to note that the division events of \textit{T. weissflogii} occur when the parent cell weighs between $310\,{\rm pg}$ and $436\,{\rm pg}$, with a mean value of $\approx 378\,{\rm pg}$. This kind of a tip-off value prediction in dry mass for a division event is achieved for the first time thanks to this method and is an example of the type of information that can be acquired by employing a non-intrusive technique that can continuously measure single cells over their life histories.

\section{Discussion}

The main advantage of combining holographic microscopy with deep learning algorithms lies in the ability to monitor position and dry mass of individual plankton cells over extended time periods. 
This method is nondestructive and minimally invasive, so that it allows quantitative assessments of trophic interactions such as feeding and biomass increase throughout the cell cycle, providing unprecedented detail to the life-histories of marine microorganisms.

The standard methods to determine the biomass of cells entail either performing elemental analysis on cells harvested from single species cultures or estimating the biomass from ``volume-to-carbon'' relationships drawn from multiple elemental analyses of similar plankton organisms of different sizes \cite{strathmann1967estimating,menden2000carbon}. 
Elemental analysis has the advantage of providing detailed measurements of individual elements, typically carbon, nitrogen, and hydrogen; however, it is destructive and cannot provide individual cell resolution. 
``volume-to-carbon'' relationships can provide biomass estimates of individual living cells as long as the volume of the cells can be measured accurately \cite{menden2000carbon}; yet, the variability around the relationship is substantial (e.g., the estimated value for \emph{Oxyrrhis marina} used here is 30\% higher than that measured by elemental analysis \cite{menden2000carbon}). 
Moreover, the ``volume-to-carbon'' relationships do not account for the nutritional status of the cell (e.g., as discussed in Results, ``Dry mass estimates'', similarly-sized cells of the same species can indeed differ by more than a factor of two in estimated dry mass by ``volume-to-carbon'' method).

Also, Fig.~\ref{figure4}f reveals asymmetric cleavage in \textit{T. weissflogii}, where  sister cells receive unequal proportions of the mother-cell biomass. 
Asymmetric division is well established in centric diatoms which have a rigid silica shell called the frustule consisting of two halves.
Both have the shape of a  cylinder with one open ended (representing a tin), and the other with a slightly smaller size (representing a lid) fitting inside the other.
Upon division the two halves part ways and both form a new hypotheca, the part that fits inside the half that is inherited from the mother cell. Hence one of the daughter-cells will be slightly smaller and the other maintains the size of the mother cell. When cells reach a critically small size, the size is reinstalled through sexual reproduction\cite{macdonald1869structure, pfitzer1869bau}.
Experimentally, asymmetric cleavage beyond the reductive cell cycle has also been shown in the diatom, \textit{Ditulum breightwelli} \cite{laney2012diatoms} where daughter cells are of different volume. Similar to in our study the sister cells show unequal times to the subsequent division (Fig.~\ref{figure4}a-e) which may lead to faster population growth\cite{laney2012diatoms}. 
Physiological asymmetry is also found in bacteria (\textit{E. coli}) where old damaged cell content is distributed differently leading to the development of age structure in prokaryote populations \cite{proenca2018age}. 
The differences in biomass of sister cells observed here is larger than expected from the volume alone which suggest that unequal division of biomass may be more common in protozoans than previously perceived.  

On the other hand, three-dimensional tracking of microorganism is not easily achieved by alternative methods. 
In many instances, two-dimensional tracks are obtained and three-dimensional swimming behavior inferred by assuming isotropic swimming \cite{selander2011grazer}. 
Moreover, traditional tracking techniques often lose track of cells when they intersect other cells or swim out of focus. Holographic microscopy overcomes the limitations of shallow depth of field in conventional light microscopy, and the three-dimensional positioning together with biomass estimates facilitates linking of cell positions into coherent trajectories. 
Three-dimensional positioning of both prey and predator cells also allows detailed observations of cell--cell interactions such as reaction distances and rejection events. 
Finally, holographic microscopy does not require manipulation such as dilution and is not sensitive to the same assumptions as the dilution technique. 
On the other hand, the volume that can be monitored by holographic microscopy is limited by the coherence length of the light source, which may result in edge effects in smaller setups. Here, we use a LED light source with a relatively short coherence length ($\approx 176\,{\rm \mu m}$ with $1\,{\rm nm}$ line filter). The depth of the observational chamber can, however, be increased by use of laser light sources with longer coherence length. 
With simple modifications to the setup design, the lensless approach used here can be adopted for smaller organisms such as bacteria and heterotrophic nanoflagellates, or larger, such as rotifers, and small crustaceans. 
It could also be merged with rotating stages that keep cells in suspension \cite{krishnamurthy2020scale}
The method is particular suitable to study micro-zooplankton grazing behaviors. 

A large and growing proportion of microphytoplankton previously considered fully autotrophic have been reclassified as mixotrophic, i.e., supporting their energy demand by a combination of photosynthesis and uptake of dissolved or particulate organic matter \cite{stoecker2017mixotrophy}. This discovery is of more than academic interest as allowing mixotrophy in food web modelling results in up to a three-fold increase in average organism size and enhanced transfer of biomass to higher trophic levels, thus increasing the sinking flux of carbon, ``the biological pump'', by an estimated 35\% \cite{ward2016marine}. 
This discovery has led to something of a paradigm shift in marine microbial ecology and highlighted the need for new methods to accurately account for mixotrophy in biogeochemical models \cite{flynn2019mixotrophic}. Mixotrophy is a plastic trait that changes with conditions. Among the more extreme cases, there is the dinoflagellate \emph{Karlodinium armiger}, which is an autotroph at low cell concentrations, but switches to heterotrophy at high densities and even revert the food web by killing their copepod predators and extract their content through peduncle feeding \cite{berge2012marine}. The combination of holographic microscopy and deep learning algorithms can be used to quantify uptake of both particulate and dissolved matter. Furthermore, it will allow us to explore the level of mixotrophy in different conditions and organisms by monitoring drymass in factorial combinations of light and organic substrates. 

We conclude that the marriage between holographic microscopy and deep learning provides a strong complementary tool in microbial ecology. 
It allows the nondestructive and minimally invasive determination of the three-dimensional position and dry mass of individual microorganisms. 
It outperforms traditional methods in terms of speed and individual resolution and rivals the precision and accuracy of current methods. While holographic microscopy has already been employed in marine sciences, the combination with deep learning algorithms makes it more versatile and an order of magnitude faster \cite{rivenson2019deep}, than holographic microscopy  without deep learning which is key to follow individuals throughout their lifespan.

\section{Methods}

{\bf Holographic imaging.}
During the measurements, the planktons are kept in enclosed circular wells (diameter $3\,{\rm mm}$, depth $1\,{\rm mm}$,, volume $\approx 10\,{\rm \mu l}$). 
As shown in Suppl. Fig.~\ref{SF1}a, the planktons within the wells are imaged using a lensless holographic imaging technique \cite{daloglu20183d}, where the sample is illuminated by a narrow-band LED light source (Thorlabs M625L3, center wavelength $632\,{\rm nm}$, bandwidth $18\,{\rm nm}$ with a $1\,{\rm nm}$ bandwidth line filter, Thorlabs FL632.8-1, centered at $632.8\,{\rm nm}$) and the sensor (Thorlabs DCC1645C, CMOS sensor area $4.608\,{\rm mm}\times 3.686\,{\rm mm}$, $1024\,{\rm px} \times 1280\,{\rm px}$) is placed immediately below the bottom of the well (the distance from the bottom of the well to the sensor is $\approx 1.5\,{\rm nm}$). 
In this way, the entire well is imaged within a single field of view (Suppl. Fig.~\ref{SF1}b), ensuring that all planktons are continuously visible for the whole duration of the experiment. 
The resulting image of the planktons are diffraction patterns formed by the interference of the unscattered light and the light scattered by the planktons. 
These diffraction patterns act as a unique fingerprint of the size, refractive index, as well as the lateral and axial position of the planktons, which has been used previously to characterize micron-scale objects \cite{altman2020catch} (See Suppl. Section 1, ``Relation between dry mass and scattering cross section''). Prior to the analysis by RU-net and WAC-net, the diffraction patterns are normalized with respect to the intensity of the unscattered light.

{\bf RU-Net architecture and training.}
To obtain the plankton positions, we use
a modified U-Net \cite{ronneberger2015u}, which we name Regression U-Net (RU-Net) implemented using DeepTrack 2.0 \cite{midtvedt2021quantitative}.
Its architecture is shown in Suppl. Fig.~\ref{SF2}c. 
The downsampling part of the RU-Net consists of a series of convolutional blocks, where each convolutional block contains a series of convolutional layers followed by a max-pooling layer and a ReLU activation. 
For an input image of size $128\,{\rm px} \times 128\,{\rm px}$ (about one-tenths the size of acquired experimental image, Fig.~\ref{SF2}a), we use 6 convolutional blocks in sequence containing 8, 16, 32, 64, 32, 32 convolutional layers, respectively. 
In the upsampling part, the RU-Net is divided into two different paths that function as two independent regular U-Nets.
Each upsampling path contains a series of 4 upsampling blocks with each containing a deconvolutional layer followed by a series of 128, 64, 32, 16 convolutional layers. 
Features obtained from each convolutional block in the downsampling path are appended to the features of the upsampling path at each upsampling block. 
One of the upsampling paths is used for the segmentation of planktons for which a sigmoid activation is applied on the output of the final upsampling block. 
The other upsampling path is used to obtain the heat maps of dry mass, axial $z$-distance, and lateral $x$- and $y$-positions (to refine the lateral localization accuracy).
Finally, the outputs of both paths are concatenated to obtain a five-channel output tensor of size $128 \times 128 \times 5$ (Suppl. Fig.~\ref{SF2}b).

To train the RU-Net, we simulate holographic images of size $128\,{\rm px} \times 128\,{\rm px}$ using DeepTrack 2.0 \cite{midtvedt2021quantitative}. 
Each image contains planktons of different sizes and refractive indices (Suppl. Fig.~\ref{SF2}a). 
Planktons are simulated in a wide dry-mass range from $1\,{\rm pg}$ to $995\,{\rm pg}$, with their corresponding equivalent spherical diameters ranging from $1.5\,{\mu m}$ to $10\, {\mu m}$. 
The RU-Net is trained using the AMSgrad optimiser \cite{reddi2019convergence}, with a learning rate of 0.0001. The model is trained on 4000 simulated holographic images in mini-batches of 16 images for 300 epochs, with a custom loss function. 
The images are generated with the continuous generator of DeepTrack 2.0 \cite{midtvedt2021quantitative} starting with 2000 images generated before the training and the remaining images generated as the training proceeds.
The training process (including the data generation) takes about 1.5 hours on a Kaggle server (Tesla P100 graphics processor unit and Intel(R) Xeon(R) CPU @ 2.00GHz).

{\bf WAC-Net architecture and training.}
To obtain a refined dry mass value, we use a weighted average convolutional neural network (WAC-Net) \cite{midtvedt2021fast} implemented using DeepTrack 2.0 \cite{midtvedt2021quantitative}.
Its architecture is shown in Suppl. Fig.~\ref{SF3}b.
The downsampling part of WAC-Net contains a time distributed block that consists of a series of convolutional blocks. Each convolutional block contains a convolutional layer followed by a ReLU activation and a max-pooling layer.
For a input image sequence consisting of frames of size $64\,{\rm px} \times 64\,{\rm px}$, we use a series of 4 convolutional blocks containing 32, 64, 128, 256 convolutional filters, respectively. 
The features are then flattened and analyzed by a series of 2 dense layers with 128 nodes to obtain the latent representations for the images in the sequence.
We use two convolutional layers with 128 and 1 filters, respectively, on each of the output latent vectors to obtain single-value representations of the weights.
These weights are further normalized with a softmax layer.
We average the latent vectors with the normalised weights to obtain a weighted representation of the latent vectors.
Finally, we use a series of dense layers with 32, 32, 2 nodes on the output representations to generate the output values of dry mass and radius.
The dry mass predicted by the WAC-Net is converted to natural mass units by using a specific refractive increment value, $\partial n/\partial c = 0.21\,{\rm ml\, g^{-1}}$ accounting for the average planktonic solute composition \cite{aas1996refractive}.

To train the WAC-Net, we simulate 15-frame sequences of $64\,{\rm px} \times 64\,{\rm px}$ images that contain a main plankton (whose dry mass and radius the WAC-Net will estimate, Suppl. Fig.~\ref{SF3}a) near the center of the image, which randomly moves by $\pm 3.6\,{\rm \mu m}$ ($\pm 1\,{\rm px}$) in the $xy$-plane and  $\pm 100\,{\rm \mu m}$ in the $z$-direction (since the frame is laterally centered on the plankton, the $xy$-plane movement is smaller than the $z$-movement).
In order to make the network robust to the existence of multiple planktons within a frame, other planktons are occasionally added to the frames. These additional planktons are given a directed in-plane motion randomly chosen between $3.6\,{\rm \mu m}$ ($1\,{\rm px}$) and $25.2\,{\rm \mu m}$ ($7\,{\rm px}$) per frame. In the $z$-direction, the motion is also randomized at $\pm 100\,{\rm \mu m}$ per frame.
The WAC-Net is trained using the AMSgrad optimiser \cite{reddi2019convergence}, with a learning rate of 0.0001.
The model is trained on 4000 images in mini-batches of 32 images for 200 epochs, with a mean absolute error (MAE) loss function. The images are generated with the continuous generator of DeepTrack 2.0 \cite{midtvedt2021quantitative}, starting with 2000 images generated before the training and with remaining images being generated as the training proceeds. 
The training process (including the data generation) takes about 45 minutes on a Kaggle server (Tesla P100 graphics processor unit and Intel(R) Xeon(R) CPU @ 2.00GHz).

{\bf Plankton cultures.}
We used a representative subset of plankton organisms covering larger primary producers such as  diatoms (\textit{Thallasiosira weissflogii}, \textit{T. pseudonana}) and dinoflagellates (\textit{Alexandrium minutum, Kryptoperidinium triquetrum}, \textit{Scripsiella acuminata}) as well as smaller flagellates (\textit{Isochrysis galbana}, \textit{Dunaliella tertiolecta}). We also included the heterotrophic dinoflagellate, \textit{Oxhyrris marina} to explore predator-prey interactions and feeding events (See Suppl. Table.~\ref{table1}, ``Planktons used in the experiments''). Plankton cultures were reared in L medium at 26 PSU salinity in a light and temperature controlled incubator (\SI{16}{\degreeCelsius}, 12h:12h light:dark cycles, $100\,{\rm fmol\, m^2 s^{-1}}$). 
The \textit{Oxyrrhis marina} cultures were fed \textit{Isochrysis galbana} or \textit{Dunaliella tertiolecta} weekly, but starved until prey cells became rare before experiments to avoid unintentional addition of prey cells to experiments.

{\bf Dry mass estimation by ``volume-to-carbon'' relationships.}
We compare the dry mass estimates from the holographic microscopy against the standard method based on ``volume-to-carbon'' relationships by measuring the volume of the cells on a Coulter counter (Beckaman, Multisizer III).
The Coulter counter is equipped with a 100-${\rm \mu m}$ orifice tube. Its accuracy is confirmed with latex beads. 
Volume estimates are subsequently used to estimate the carbon content of the cells using the equations given in Ref.~\cite{menden2000carbon}.

{\bf Dry mass estimation by elemental analysis.}
A precise algal culture volume of known cell concentration is filtered onto pre-combusted  ($450^\circ{\rm C}$) $25\,{\rm mm}$ glass fiber filters (Whatman GF/F). 
The filters are dried overnight at $60^\circ{\rm C}$. 
Carbonates are removed by incubation in an exicator together with fuming hydrochloric acid. 
The filters are enclosed in tin capsules and analysed on an elemental analyzer (ANCASL, SerCon, UK) coupled to an isotope ratio mass spectrometer (20–20, SerCon, UK).

\providecommand{\noopsort}[1]{}\providecommand{\singleletter}[1]{#1}%

\vspace{5mm}
\noindent {\bf Competing interests}\\
The authors declare no competing interests.\\\\
\noindent {\bf Code availability}\\
Data and code are available upon reasonable request.\\\\
\noindent {\bf Author contributions}
HB, DM, ES, GV conceptualized the work.
HB and DM designed the method, carried out the simulations and trained the neural networks. 
HB and ES performed the experiments. 
HB analyzed the data and generated the figures.  
DM and BM implemented the neural network architectures. 
HB, DM, ES and GV interpreted the results and wrote the manuscript with inputs from all the authors. 
ES and GV obtained the funding. 
DM, ES and GV supervised the project.
\\\\
\noindent {\bf Acknowledgements}\\
The authors would like to thank Jan Heuschele for the illustrations, and Olga Kourtchenko for providing plankton cultures. This work was partly supported by the H2020 European Research Council (ERC) Starting Grant ComplexSwimmers (Grant No. 677511), the Horizon Europe ERC Consolidator Grant MAPEI (Grant No. 101001267), the Knut and Alice Wallenberg Foundation (Grant No. 2019.0079), and the Swedish Research Council (VR, Grant No. 2019-05238)


\clearpage
\pagebreak
\widetext
\begin{center}
\section{Supplementary Materials}
\end{center}
\setcounter{figure}{0}
\renewcommand{\thefigure}{S\arabic{figure}}
\renewcommand{\theHfigure}{S{\arabic{figure}}}

\subsection{Supplementary Figures}
\begin{figure*}[h]
    \includegraphics[scale=0.8, angle=0]{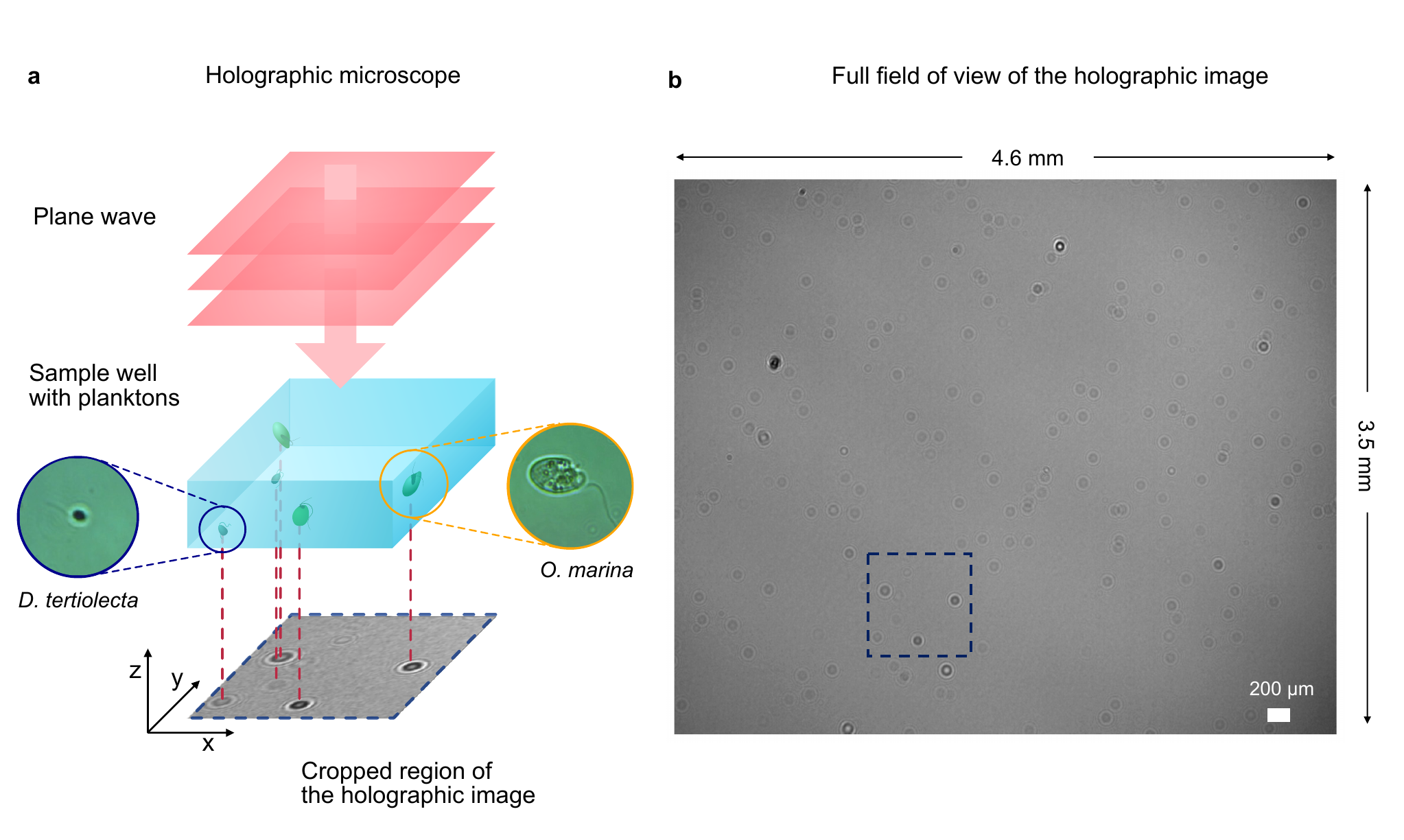}
    \caption{
    \label{SF1} 
    \textbf{Holographic microscope and full-scale view of an experimental holographic image.}
    {\bf a} Detailed view of the inline holographic microscope shown in Fig.~\ref{figure1}a: A plane wave generated by a monochromatic LED source illuminates the sample well that contains two different species of planktons ({\it Dunaliella tertiolecta} and {\it Oxyrrhis marina}), whose bright-field microscope images are shown in the insets. The holograms of the planktons are shown in the cropped holographic image at the bottom.
    {\bf b} Full-scale hologram: Inline holographic microscopes offer a large field of view; in our case, the active region of the sensor has the dimensions $4.6\,{\rm mm} \times 3.5\,{\rm mm}$. The cropped region in Fig.~\ref{figure1}a is marked by the dashed box.}
\end{figure*}

\clearpage
\begin{figure*}[h]
    \includegraphics[scale=0.9, angle=0]{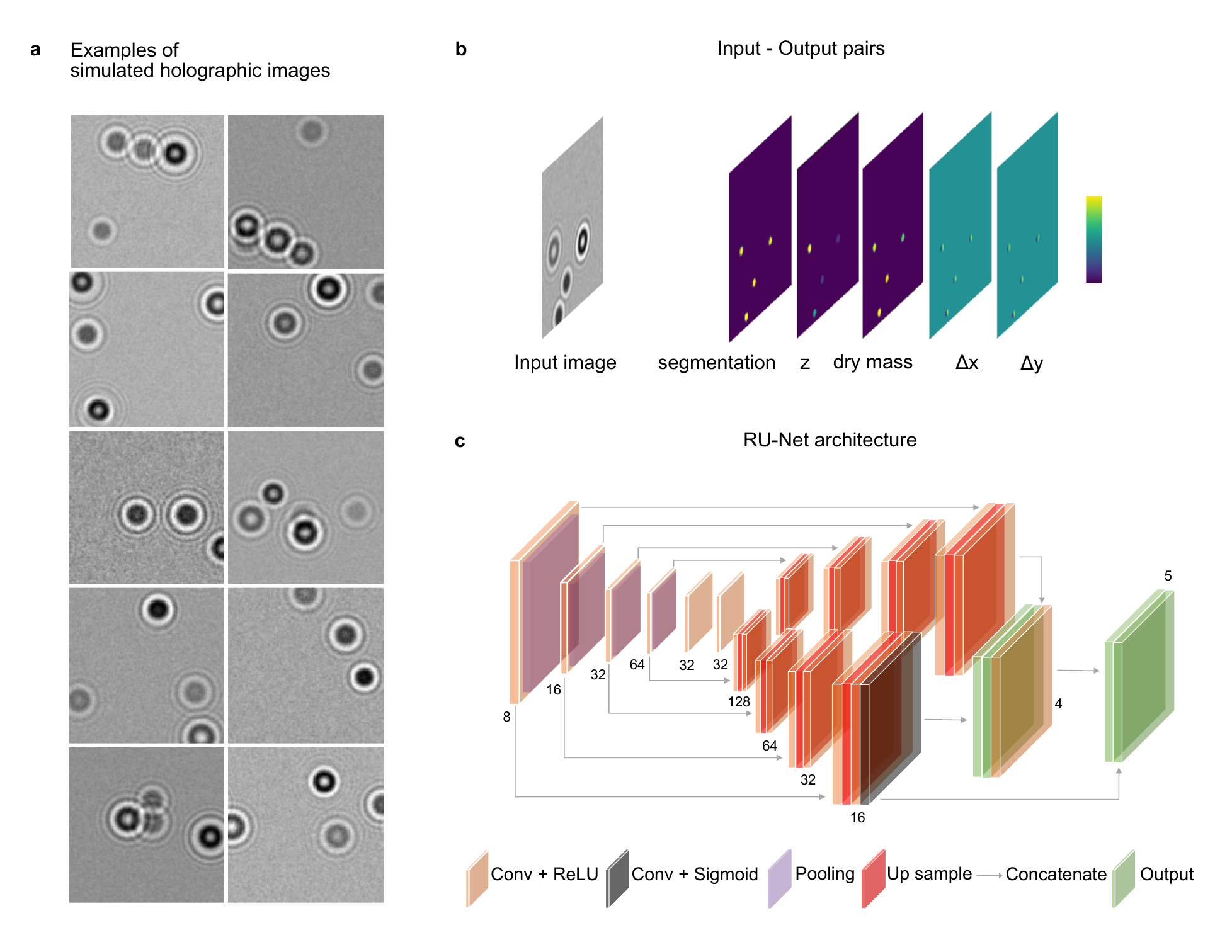}
    \caption{
    \label{SF2}
    {\bf Simulated holographic images and RU-Net architecture.}
    {\bf a} Examples of simulated plankton holograms generated by DeepTrack 2.0 \cite{midtvedt2021quantitative}: Each image ($128\,{\rm px} \times 128\,{\rm px}$) contains plankton holograms generated over a wide range of sizes and refractive indices. The background noise is varied randomly with a Gaussian random noise to match the experimental images.
    {\bf b} Input--output pairs to train the RU-Net: The input of the RU-Net is a simulated holographic image. The output maps contain the segmentation of the planktons, their $z$-position, their dry mass $m$, and the distances $\Delta x$ and $\Delta y$ from the closest plankton for each pixel (to be used for the accurate localization of planktons).
    {\bf c} RU-Net architecture: The downsampling part consists of a sequence of 6 convolutional blocks containing a convolutional layer (depicted in brown; the number of convolutional filters is indicated at the bottom of each convolutional layer) followed by a ReLU activation and a max-pooling layer (depicted in purple). In the upsampling path, the RU-Net is divided into two branches. Each upsampling branch contains a sequence of 4 upsampling blocks, where each upsampling block contains an upsampling layer (depicted in red) followed by a convolutional block (depicted in blown), and a ReLU activation. The final upsampling block used for the plankton segmentation employs a sigmoid activation (depicted in black). The outputs of both upsampling paths (depicted in green) are merged together, after which a final convolutional block is added to predict the five-channel output image. 
    See also Methods, ``RU-Net architecture and training''.
    } 
\end{figure*}
\clearpage

\begin{figure*}[h]
    \includegraphics[scale=0.89, angle=0]{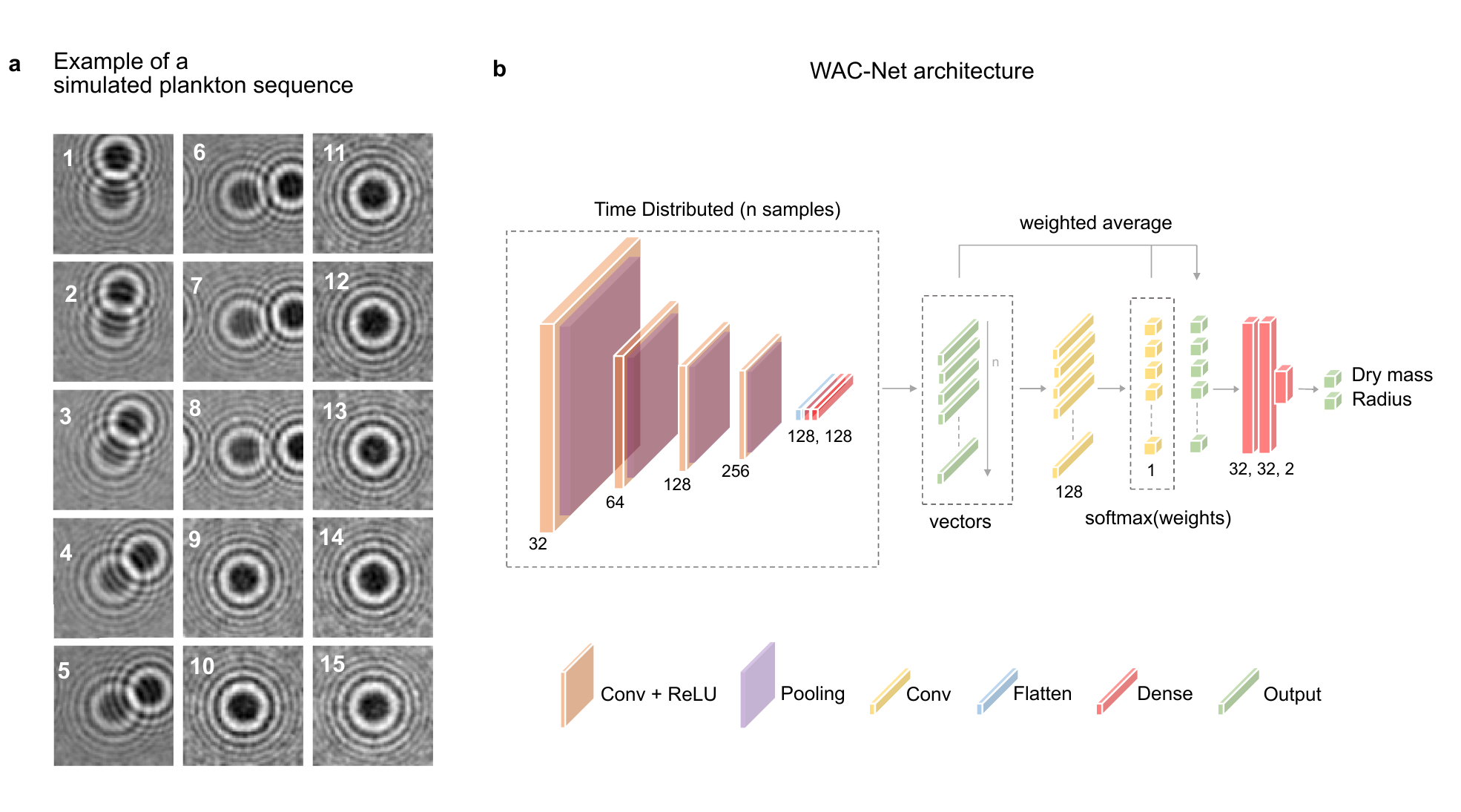}
    \caption{
    \label{SF3}
    {\bf Simulated plankton sequence and WAC-Net architecture.}
    {\bf a} Simulated time sequence of holographic images cropped around an individual plankton: The 15-frame simulated sequence is an example input sequence for training the WAC-Net to estimate the plankton dry mass and radius. The images are centered on the target plankton, while other planktons in the background are considered as noise that the WAC-Net should learn to ignore.
    {\bf b} WAC-Net architecture: The input image sequence is processed by a time-distributed block consisting of a series of 4 convolutional blocks. Each convolutional block has a convolutional layer (depicted in brown; the number of convolutional filters is indicated at the bottom of each convolutional layer), followed by a ReLU activation and pooling layer (depicted in purple). The features are flattened (depicted in blue) followed by a series of 2 dense layers (depicted in red; the number of nodes is indicated at the bottom), which return an output vector (depicted in green) for each image image in the sequence ($n = 15$ images). This is followed by two convolutional layers (depicted in yellow; with 128 and 1 filters, respectively), which estimate the weight for each image in the sequence. A weighted average is then obtained for the vectors to obtain $n$ single value outputs (depicted in green). Finally, the outputs are processed with a series of dense layers to obtain the final result for the dry mass and radius. 
    See also Methods, ``WAC-Net architecture and training''.) 
    }
\end{figure*}
\clearpage

\begin{figure*}[h]
    \centering
    \includegraphics[scale=0.86, angle=0]{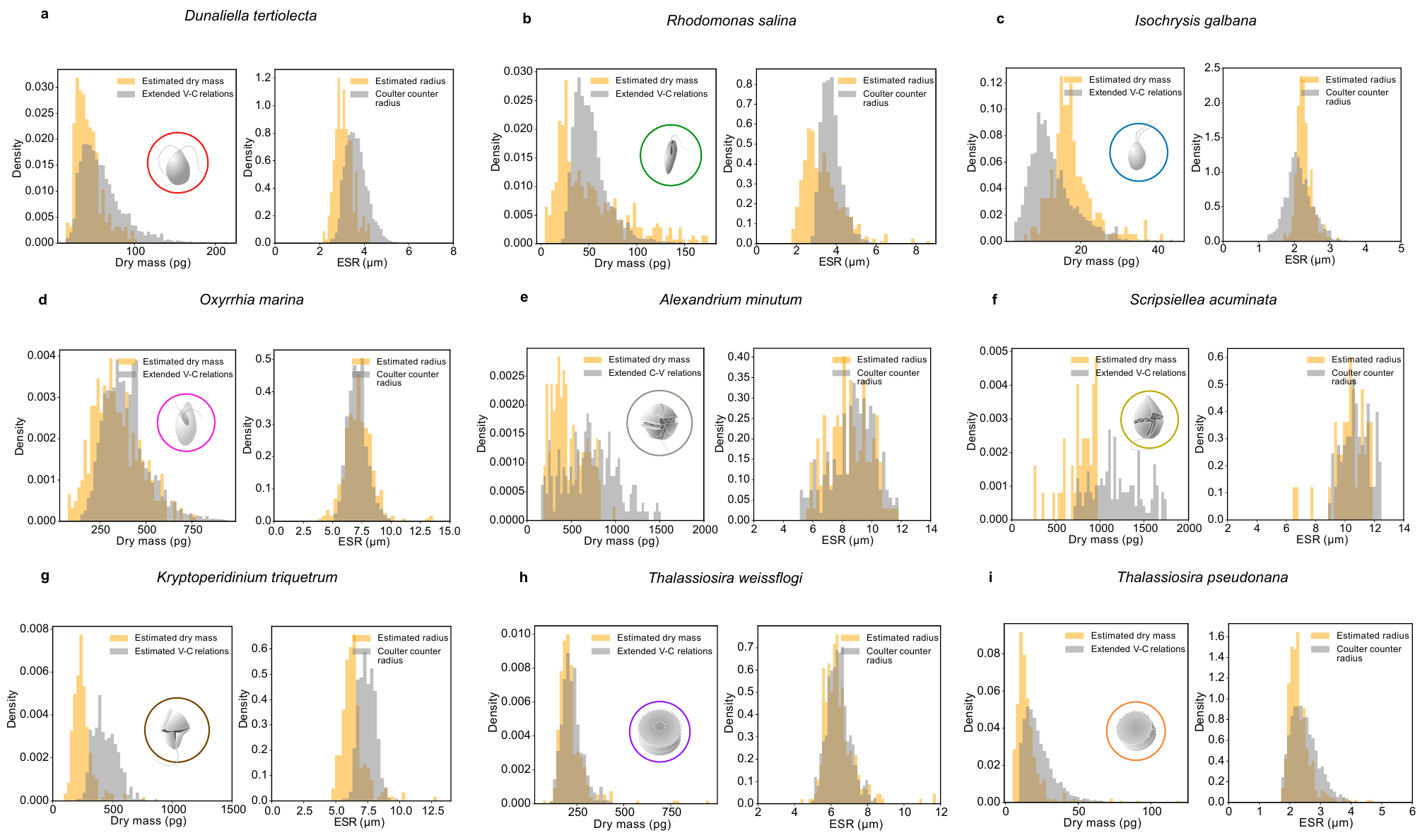}
    \caption{
    \label{SF4} 
    {\bf Dry mass and equivalent spherical radius (ESR) estimates for different species of planktons.}
    Comparison between dry mass (left panels) and ESR (right panels) estimated from ``volume-to-carbon'' relationships \cite{menden2000carbon} (gray histograms) and dry mass estimated using our method (orange histograms) for 9 species of plankton.
    The plankton species include, phytoplankton species ({\bf a} {\it D. tertiolecta}, {\bf b} {\it R. salina}, {\bf c} {\it I. galbana}), microzooplankton species ({\bf d} {\it O. marina}, {\bf e} {\it A. minutum}, {\bf f} {\it S. acuminata}, {\bf g} {\it K. triquetrum}), and diatoms ({\bf h} {\it T. weissflogii}, {\bf i} {\it T. pseudonana}).}
\end{figure*}
 
\begin{figure*}[h]
    \includegraphics[scale=0.85, angle=0]{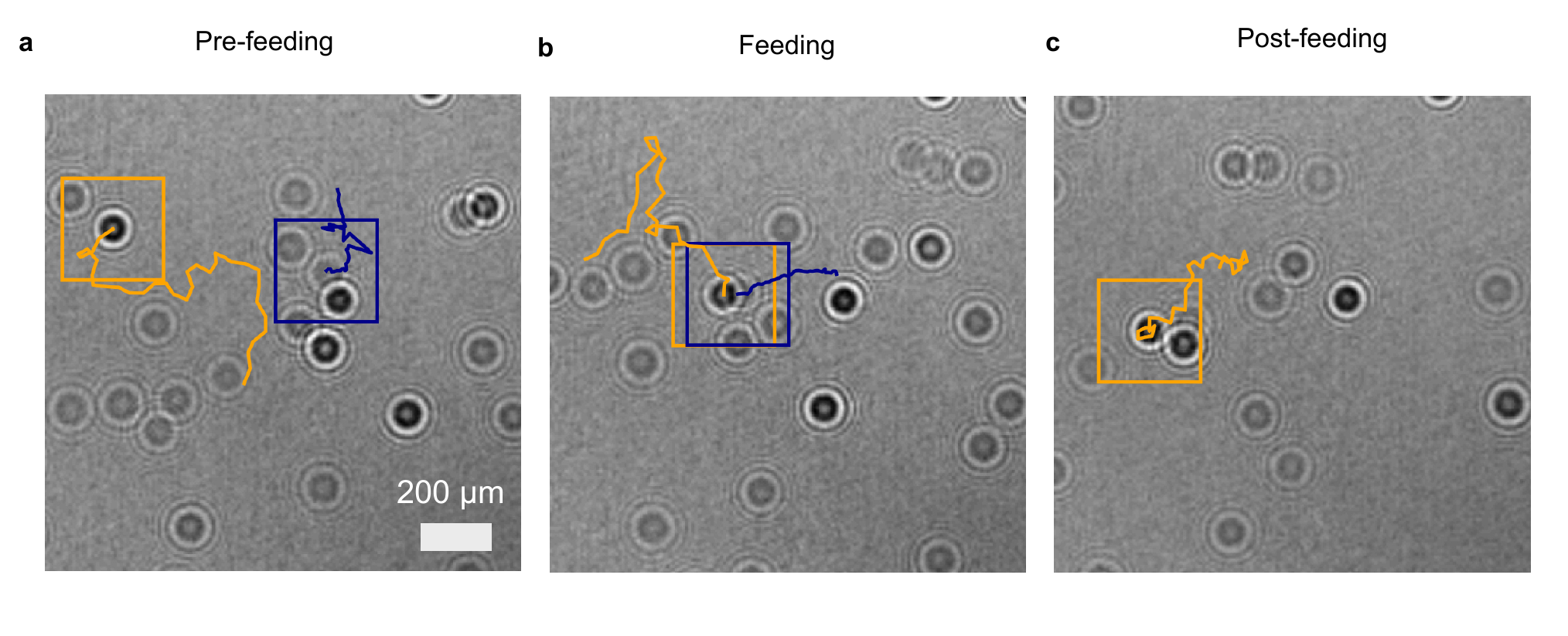}
    \caption{
    \label{SF5} 
    {\bf 2D projection of traces in a feeding event.}
    {\bf a-c} 2D recording of a feeding event where 
    {\bf a} a predator micro-zooplankton ({\it Oxyrrhis marina}, orange traces) approaches a prey phytoplankton ({\it Dunaliella tertiolecta, blue traces}), 
    {\bf b} feeds on it, and 
    {\bf c} finally moves away (see Figs.~\ref{figure3}a-c). 
    The trajectories show the path of planktons in last 32 frames; the current position is highlighted with a rectangle (orange for predator and blue for prey).
    }
\end{figure*}
\clearpage

\subsection{Supplementary Videos}
\setcounter{figure}{0}
\renewcommand{\thefigure}{V\arabic{figure}}
\renewcommand{\theHfigure}{V{\arabic{figure}}}

We provide supplementary videos corresponding to the feeding event and cell-division event discussed in the articles.
\par\null\par
{\bf Supplementary Video 1 --- Feeding event main}

\begin{figure*}[h]
    \includegraphics[scale=0.88, angle=0]{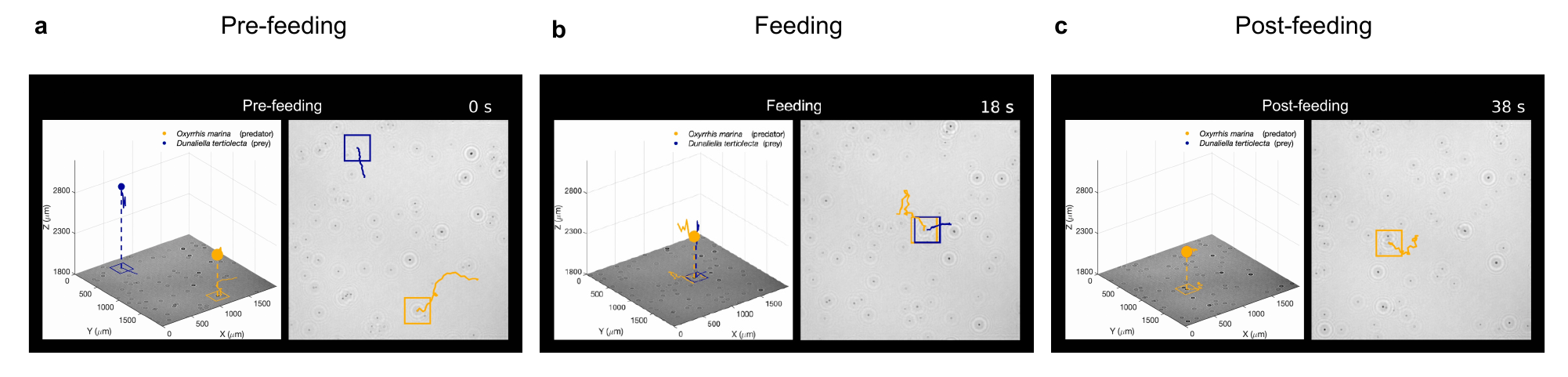}
    \caption{
    \label{SV1} 
    {\bf Feeding event.}
    {\bf a-c} Snapshots of supplementary video showcasing the feeding event discussed in Fig.~\ref{figure3}: 3D motion of the prey ({\it D. tertiolecta}, blue) and predator ({\it O. marina}, orange) with holographic image reproduced on the bottom of the 3D volume on the left.
    Their 2D motion at a reconstructed holographic plane is shown on the right. The choice of the reconstructed plane (at $z = 2400 \, {\rm \mu m}$ from the sensor) ensures a clear view of the most of the planktons.
    The feeding phase and experimental time are indicated at the top of the video.
    }
\end{figure*}

{\bf Supplementary Video 2 --- Feeding event additional}

\begin{figure*}[h]
    \includegraphics[scale=0.88, angle=0]{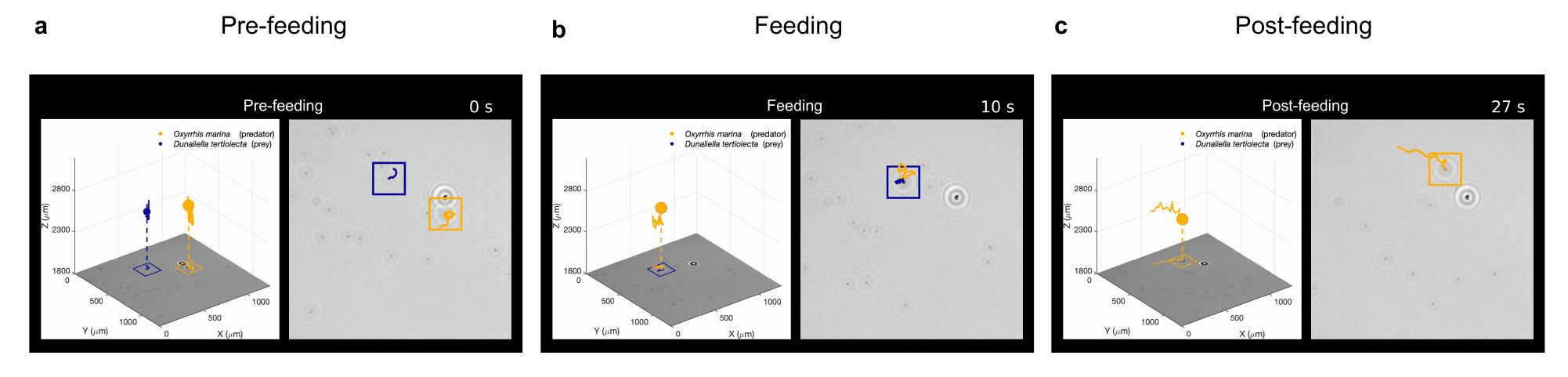}
    \caption{
    \label{SV2} 
    {\bf Additional feeding event.}
    Snapshots of supplementary video showcasing an additional feeding event.
    }
\end{figure*}

{\bf Supplementary Video 3 --- Division event}

\begin{figure*}[h]
    \includegraphics[scale=0.88, angle=0]{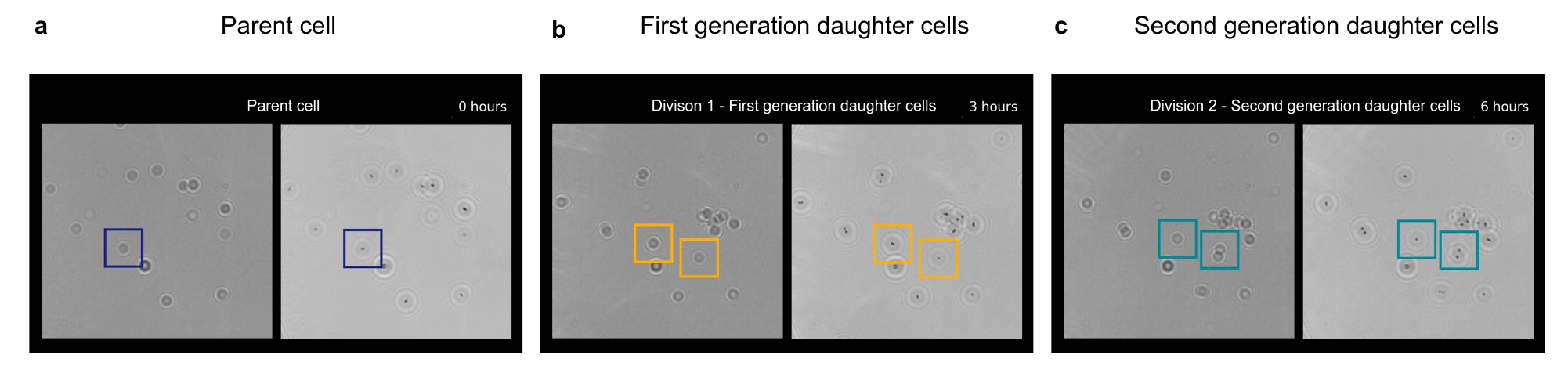}
    \caption{
    \label{SV3} 
    {\bf Division event.}
    Snapshots of supplementary video showcasing the life history of plankton in Fig.~\ref{figure4}. 
    The holographic image of a diatom ({\it T. weissflogii}) undergoing successive cell division events is shown by the holographic image on the left. 
    The image on the right is the reconstructed image on the plane $z = 2400 \, {\rm \mu m}$ as before. 
    The generation and experimental time are indicated at the top of the video.
    }
\end{figure*}

\clearpage

\vspace{5mm}
\begin{center}
\setlength{\tabcolsep}{0.1cm}
\renewcommand{\arraystretch}{1.5}
\begin{table}
\begin{tabularx}{0.8\textwidth}{
    |>{\centering\arraybackslash}X
    |>{\centering\arraybackslash}X
    |>{\centering\arraybackslash}X
    |>{\centering\arraybackslash}X | }
    \hline
    {\bf Scientific name} & {\bf Strain identifier} & {\bf Class} & {\bf ESD (mean $\pm$ sd)}\\
    \hline
    {\it Alexandrium minutum} & GUMACC83 (CCMP113) & Dinophyceae & $18.3 \pm 2.5\,{\rm \mu m}$\\
    \hline
    {\it Dunaliella tertiolecta} & GUMACC5 & Cholorphceae & $6.7 \pm 0.9\,{\rm \mu m}$\\
    \hline
    {\it Isochrysis galbana} & GUMACC108 (CCMP1323) & Prymnesiophyceae & $4.0 \pm 0.7\,{\rm \mu m}$ \\
    \hline
    {\it Kryptoperidinium triquetrum} & GUMACC71 (LAC20, KA86) & Dinophyceae & $14.9 \pm 1.2\,{\rm \mu m}$\\
    \hline
    {\it Oxyrrhis marina} & DTU-Aqua & Dinophyceae & $14.5 \pm 1.5\,{\rm \mu m}$\\
    \hline
    {\it Rhodomonas salina} & GUMACC126 (CCAP978/27) & Cryptophyceae & $7.4 \pm 1.0\,{\rm \mu m}$\\
    \hline
    {\it Scripsiella acuminata} & GUMACC110 (CCMP1331) & Dinophyceae & $17.7 \pm 4.9\,{\rm \mu m}$\\
    \hline
    {\it Thallassiosira pseudonana} & GUMACC132 (CCAP1085/12) & Cosconodiscophyceae & $4.8 \pm 0.9\,{\rm \mu m}$\\
    \hline
    {\it Thallassiosira (Conticribra) weissflogii} & GUMACC162 (CCAP1085/18) & Cosconodiscophyceae & $12.9 \pm 1.4\,{\rm \mu m}$ \\ 
    \hline

\end{tabularx}
\caption{{\bf Planktons used in the experiments.} Strain identifier denotes strain code in Gothenburg University Marine Algae Culture Collection (GUMACC) and synonym strain identifier in parenthesis. The {\it Oxyrrhis marina} culture was kindly provided by Denmark Technical University (DTU-Aqua) and does not have a strain ID. Equivalent Spherical Diameter (ESD) denotes the spherical diameter based on Coulter counts (Beckman multisizer III) of pure cultures.}
\label{table1}
\end{table}
\end{center}

\clearpage

\subsection{Supplementary Section 1 --- Relation between dry mass and scattering cross section}

The amount of light scattered in any given direction from an object is a function of the shape and the refractive index of the object. The scattering amplitude $S(\theta)$ quantifies the amount of light that is deflected an angle $\theta$ from the direction of the illuminating light. For objects significantly larger than the illuminating wavelength ($2\pi R/\lambda\gg 1$, where $R$ is the radius of the object and $\lambda$ is the wavelength of the light) and small refractive index difference ($|n_{\rm object}/n_{\rm med}-1|\ll 1$, where $n_{\rm object}$ is the refractive index of the object and $n_{\rm med}$ is that of the medium), the scattering amplitude in the far field region can be estimated within the anomalous diffraction approximation as \cite{streekstra1993light}
\begin{equation}
    S(\beta,\gamma) = \left(k^2/2\pi\right)\int\int\left[1-\exp(-i \phi(\epsilon,\nu))\right]
    \times\exp(i k (\epsilon\beta + \nu \gamma) ) d\epsilon  d\nu,
\end{equation}
where $k$ is the light wave number, $\epsilon,\, \nu$ are the in-plane spatial coordinates at the axial plane of the object (the plankton), while $\beta, \, \gamma$ are the coordinates at the axial plane of the sensor, scaled by the distance $r$ between plankton and sensor position, $\beta=x/r$ and $\gamma=y/r$ with $r=(x^2+y^2+z^2)^{1/2}$. Parametrizing the scattering coordinates in spherical coordinates, one has that $\beta=\sin\theta \cos\varphi$, $\gamma=\sin\theta \sin\varphi$. Averaging the scattering amplitude over the angle $\varphi$ one obtains
\begin{equation}
    S(\theta) = k^2 \int\int\left[1-\exp(-i \phi(\rho,\chi))\right]
    \times J_0(k \rho\sin\theta)\rho d\rho d\chi,
\end{equation}
with $\rho=\sqrt{\epsilon^2+\nu^2}$ and $\sin\chi=\epsilon/\rho$. Since the planktons in our setup are measured in transmission, the signal relates primarily to small angle scattering. To lowest order in the scattering angle one obtains
\begin{equation}
    S(\theta)\approx S(0)\approx k^2 A\left(1-\cos\langle \phi\rangle - i \sin\langle \phi\rangle\right),
\end{equation}
where the spatial distribution of the phase shift over the particle is replaced by its average value, denoted by $\langle\cdot \rangle$, and $A$ is the cross sectional area of the object. The average value of the phase shift can be can be directly related to the dry mass of the particle by noting that the local phase shift due to a weakly scattering object with refractive index $n$ is given by
\begin{equation}
    \phi(\epsilon,\nu) = Z(\epsilon,\nu) (n-n_{\rm med}),
\end{equation}
where $Z(\epsilon,\nu)$ is the local thickness of the object and $n_{\rm med}$ is the refractive index of the solution. The refractive index of a biological object is generally linearly related to the concentration $c$ of bio-molecules within the object \cite{zangle2014live}, so that $n=n_{\rm med}+ c\cdot \frac{dn}{dc}$. As a consequence, one has that
\begin{equation}
\label{Eq:phase}
    \langle \phi \rangle = \langle kZ\rangle \cdot (n-n_{\rm med}) = kV/A\cdot c\frac{dn}{dc} =km/A\cdot\frac{dn}{dc},
\end{equation}
where $m$ is the mass of the biomolecules within the object, $\langle Z\rangle$ is the average thickness of the object, $V$ is the volume of the object and $A$ is the cross sectional area of the object. Thus, the scattering amplitude depends explicitly on the mass of biological objects. 

In order to relate the scattering amplitude to the measured intensity of the scattering pattern, we note that the intensity is given by 
\begin{equation}
    I=|E_0|^2\left|1-i\frac{S}{kr}\exp(i\,k(r-z))\right|,
\end{equation}
where $E_0$ is the amplitude of the incoming light.

From this expression, the change in light intensity in the middle of the scattering pattern compared to the unscattered light, normalized by the intensity of the incoming light, is given by

\begin{equation}
    \Delta I(0) = \frac{I}{|E_0|^2}-1 = \left(2/kz\right)\Im S =\left(2kA/z\right) \sin\left[km/A \left(dn/dc\right)\right]
\end{equation}

To lowest order in the phase shift $\phi$, the scattering intensity at $0$ degree scattering angle therefore directly proportional to the mass of a biological object. 

This formalism provides a direct means to assess the influence of absorption on the estimated mass. Light absorption is quantified by the imaginary part of the refractive index. Adding an imaginary part to the refractive index in equation \eqref{Eq:phase}, such that $n=n_0 + i\eta$, the light intensity in the forward direction becomes

\begin{equation}
    \Delta I(0) = \left(2kA/z\right) \sin\left[km/A \left(dn/dc\right)\right]\exp\left(-k\eta \langle Z\rangle \right)
\end{equation}

Using $\eta=0.002$ as a typical value for phytoplankton \cite{qi2016determination}, and an average thickness of $5{\rm \mu m}$, the correction due to absorption would amount to about $10\%$. 

\end{document}